\documentclass{JHEP}

\usepackage{graphicx}
\usepackage{psfrag}

\providecommand{\MSbar }{\ensuremath{ \overline{\rm MS} }}
\providecommand{\MSbarbold }{\ensuremath{ \overline{\rm \bf MS} }}

\makeatletter 
\preprint{hep-ph/0001040\\
\@date
}
\makeatother

\title{Subtraction method for NLO corrections in Monte-Carlo 
       event generators for leptoproduction}
\author{John Collins%
        \\
        Physics Department,
        Penn State University, \\
        104 Davey Laboratory,
        University Park PA 16802,
        U.S.A.
\\ \email{collins@phys.psu.edu}
}

\date{7 August 2000}

\keywords{QCD, NLO Computations, Deep Inelastic Scattering}

\abstract{%
In the case of the gluon-fusion process in deep-inelastic
leptoproduction, I explicitly show how to incorporate NLO corrections
in a Monte-Carlo event generator by a subtraction method.
I calculate the parton densities to be used by the event generator in
terms of \MSbar{} densities.
The method is generalizable.
A particular motivation for treating the gluon-fusion process is to
treat diffractive deep-inelastic scattering properly, since in
diffractive scattering the gluon density dominates the quark
densities. 
I also propose a modified algorithm for treating parton kinematics in 
event generators; the new algorithm results in much simpler formulae
for the NLO corrections.
A disadvantage of the new method is that some of the generated events
may have negative weights.
However, an adjustable cut-off function is present in the formalism,
and this permits a renormalization-group-like transformation that can
be used to at least reduce the proportion of events with negative
weights. 
}

\begin{document}

\section{Introduction}
\label{sec:intro}

In an event generator for leptoproduction\footnote{
   Of course, exactly similar considerations apply to event generators
   for other processes. 
}
one has a choice of generating events either by using the lowest-order
parton model process or using the next-to-leading order (NLO) hard
scattering matrix elements (for photon-gluon fusion, etc).  The first
case is suitable for the total DIS cross section, where one neglects
the NLO subprocesses, since they represent order $\alpha_s$ corrections to
the basic process.  The second case is suitable, for example, when
two-jet production is to be calculated.

Ideally, one wants to include both the LO and the NLO terms to get
good accuracy.  The problem is particularly acute in {\em diffractive} 
DIS, since the gluon density is substantially larger than the
quark density.  Thus the gluon-induced NLO subprocess is not
necessarily smaller than the LO parton model process.  This implies that
inclusion of the photon-gluon term is mandatory to get a sensible
phenomenology.  However, the initial-state showering associated with
the LO process already includes part of the photon-gluon process, and
it is not at all obvious how the two terms are to be combined.

In this paper, I explain how to consistently use both terms in an
event generator like LEPTO \cite{LEPTO} or RAPGAP \cite{RAPGAP} that
uses the algorithm constructed by Bengtsson and Sj{\"o}strand \cite{BS}
for initial-state showering.  As a consequence of their method for
treating the exact parton kinematics, the resulting formulae have a
non-linear dependence on the parton densities.  Therefore, I also
propose an alternative leading-order algorithm for which the
corresponding NLO corrections have a conventional structure.

With the exception of the recent paper by Friberg and Sj{\"o}strand
\cite{FS}, previous attempts, e.g., \cite{MC.NLO,mrenna}, 
at incorporating NLO corrections have tended to implement them by a
reweighting of the events generated by showering from the LO matrix
elements.  In contrast, the subtraction method that I describe
involves generating two classes of events.  One class is made from the
LO parton-model process by showering the initial and final state
quarks, exactly as at present.  The second class of events is
generated by starting with a photon-gluon fusion process, and
showering the partons, again exactly as at present, but with one
exception.  The exception is that the hard cross section for the
photon-gluon fusion subprocess is equipped with a subtraction that
correctly compensates the double counting between the two classes of
events; the subtraction removes that part of the photon-gluon fusion
term that is included in the LO parton model plus showering
calculation.  My method is rather similar to the one outlined by
Friberg and Sj{\"o}strand in \cite{FS}.

Of course, there is no physical distinction between the two classes of
events; they populate the same regions of state space, and the events
differ only in how the program generates them.  Indeed the relative
contributions of the two classes of events can be changed by changing
the cut-off on the virtuality of lines in the parton showering.  A
change of this cut-off amounts to a renormalization-group
transformation, and does not affect the physical cross section, aside
from the error due to uncalculated higher order corrections.

There is a contrast with the standard subtraction method used in
analytic calculations.  In that method an infinite number of positive
weighted events is generated from the basic photon-gluon fusion
graphs. The subtraction term gives an infinite number of events with
negative weights and somewhat different kinematics.  Although the
infinities cancel in the integral over the hadronic final states, the
integrand is totally unsuitable for use in an event generator.  The
new method does the subtractions point-by-point in the integrand, so
that a Monte-Carlo integration is entirely satisfactory.

It is true that a finite number of events with finite negative weights
may be generated with my algorithm.  This is at least a potential
disadvantage, particularly if it is desired to compare the results
directly with data.  I will discuss this problem in more detail in
Sec.\ \ref{sec:negative-weights}.  There I will show how to overcome
this problem, to a large extent, by adjustment of a cut-off function
that is present in any Monte-Carlo algorithm.  I will also explain
that if an event generator gives a small number of negative weighted
events, this is not necessarily a severe disadvantage.

Since an important immediate application is to deal with the problem
of the large gluon density in diffractive DIS, I give results for the
photon-gluon fusion process.  This case is technically simple because
there are no soft gluon effects and no virtual graphs.  However, the
method is capable of being generalized.

In Sec.\ \ref{sec:algorithm}, I summarize the basic algorithm used in
the event generator at LO, and then compute the corresponding
first-order term for the photon-gluon fusion process.  Then in Sec.\
\ref{sec:correction}, I explain the observation of Bengtsson and
Sj{\"o}strand \cite{BS} that the intended kinematics for the target hadron
are not correctly reproduced.  After reviewing the implementation of
their correction to the kinematics, I obtain the corrected first-order
cross section, from which I compute the subtracted gluon-fusion cross
section.  This forms the primary quantitative result of this paper: it
is intended to be directly implemented in an event generator.  
The parton densities are not in the
\MSbar{} scheme, so I show how to relate them to the ordinary \MSbar{}
parton densities.  Unfortunately the resulting formulae are rather
complicated; in fact, they are non-linear functionals of the parton
densities.  So in Sec.\
\ref{sec:new.algorithm}, I present a new algorithm for treating the parton
kinematics and show that the resulting NLO corrections are simpler and
more conventional than with the Bengtsson-Sj{\"o}strand algorithm.
In Sec.\ \ref{sec:negative-weights}, I discuss the problem that events
with negative weights may be generated, and show how they may be at
least reduced in number.
Finally, I summarize the directions for future work in Sec.\
\ref{sec:future}.

\section{Basic Monte-Carlo algorithm}
\label{sec:algorithm}

\subsection{Algorithm}

I first review the algorithm \cite{BS} used in LEPTO or RAPGAP.  Only
the first part of the algorithm will be relevant for our later
discussions:\footnote{
   I describe the algorithm for the case of fully inclusive DIS.  The
   case of diffractive DIS is handled by changing the proton to a
   Pomeron, by replacing the variable $x$ by $\beta$, etc.
}
\begin{enumerate}

\item Generate values of $x$ and $y$ (and hence $Q$) from the LO cross 
  section for DIS:
  \begin{equation}
  \label{LO}
     \frac{d\sigma}{dx\,dy}
     = K F_2(x,Q^2),
  \end{equation}
  with
  \begin{equation}
     K = \frac{4\pi \alpha_{\rm em}^2}{sx^2y^2}
       \left( 1-y+y^2/2 \right),  
  \end{equation}
  and 
  \begin{equation}
  \label{F2.LO}
     F_2 = \sum_a e_a^2 x f_a(x,Q^2).
  \end{equation}
  Here, the sum is over all flavors of quarks and antiquarks, and $f_a$
  is the parton probability density. 

\item 
   \label{Q1}
   Generate a virtuality $Q_1^2$ for the incoming quark $a$, a
   longitudinal momentum fraction $z_1$ for the first branching, and
   an azimuthal angle $\phi$ for this branching. The distributions arise 
   from the Sudakov form factor
   \begin{equation}
   \label{Sudakov}
      S_a(x, Q_{\rm max}^2, Q_1^2) 
     =
     \exp\left\{
         -\int_{Q_1^2}^{Q_{\rm max}^2} \frac{dQ'^2}{Q'^2}
          \frac{\alpha_s(Q'^2)}{2\pi}
          \sum_c \int_x^1 \frac{dz}{z} \, P_{c\to ab}(z) 
                        \frac{f_c(x/z,Q'^2)}{f_a(x,Q'^2)}
      \right\}.
   \end{equation}
   Here $Q_{\rm max}^2$ is normally set equal to $Q^2$. 
   The Sudakov form factor is the probability that the virtuality of
   the struck quark is less than $Q_1^2$.

\item Iterate the branching for all initial-state and final-state
   partons until no further branchings are possible.

\item Generate 4-vectors for the momenta of all the generated
   partons. 

\end{enumerate}

\subsection{First-order term in Monte-Carlo}
\label{sec:MC1}

Our aim in this paper is to calculate the NLO contribution to
deep-inelastic scattering from gluon-fusion graphs like Fig.\
\ref{fig:gamma-gluon}.  The result is to be accurate for the case that 
the incoming gluon, of momentum $p_3$, has virtuality and transverse
momentum small compared with $Q$, and that the intermediate quark, of
momentum $p_1$, has a virtuality of order $Q^2$.  To avoid double
counting, it is necessary to subtract the contribution in this region
that is obtained from the showering algorithm applied to the initial-
and final-state partons of the LO partonic cross section.

\FIGURE{
\centering
\psfrag{q}{$q$}
\psfrag{p_1}{$p_1$}
\psfrag{p_{1'}}{$p_{1'}$}
\psfrag{p_2}{$p_2$}
\psfrag{p_3}{$p_3$}
\includegraphics[scale=0.5]{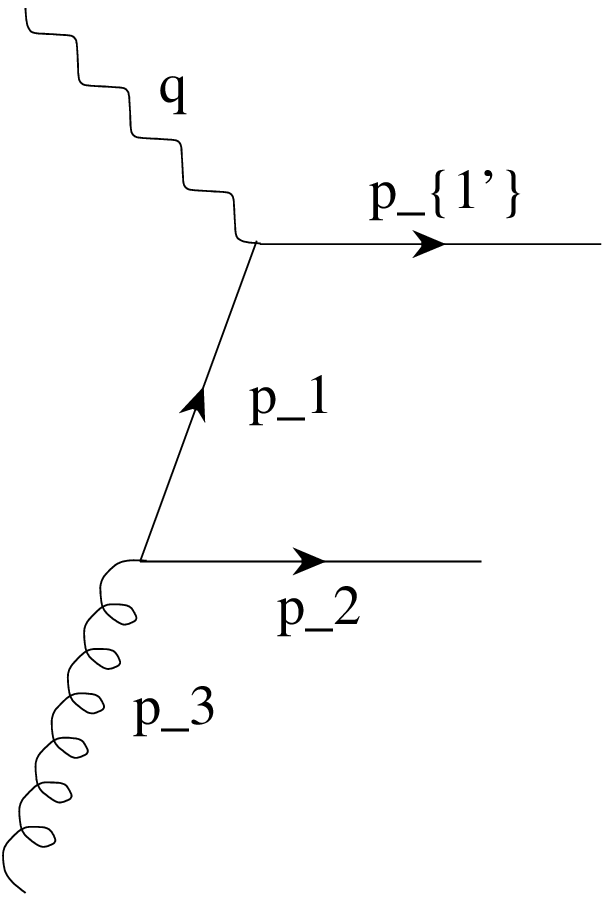}
\caption{Photon-gluon fusion.}
\label{fig:gamma-gluon}
}

This contribution is obtained by multiplying the lowest order cross
section, from Eqs.\ (\ref{LO}--\ref{F2.LO}), by the the first order
term in the expansion of the Sudakov form factor (\ref{Sudakov}) in
powers of $\alpha_s(Q^2)$.  This is made differential in the momenta of
the particles involved, and the gluon-induced term is selected:
\begin{equation}
\label{Collinear.gamma.g}
     \frac{d\sigma}{dx \, dy \, dQ_1^2 \, dz_1 \, d\phi}
  = K  \sum_a e_a^2 
    \frac{\alpha_s(Q^2)}{4\pi^2 Q_1^2}
       C(Q_1^2)
       P(z_1) \frac{ x }{ z_1 }  f_g(x/z_1, Q^2) .
\end{equation}
Here the splitting kernel is for
$g\to\mbox{quark}+\mbox{antiquark}$: $P(z_1) = P_{g\to q\bar q}(z_1) =
\frac12 (1-2z_1+2z_1^2)$.  Note that because we are doing a strict
expansion in powers of $\alpha_s(Q^2)$, the scale argument of the
parton density is $Q^2$.  The function $C(Q_1^2)$ is a cut-off
function that gives the maximum value of $Q_1^2$.  To reproduce the
upper limit on the $Q'^2$ integral in the Sudakov form factor, Eq.\
(\ref{Sudakov}), one sets $C(Q_1^2) = \theta(Q^2-Q_1^2)$.  However, as
explained in the introduction, one can envisage changing the cut-off
function.

One might worry that a full Sudakov form factor should appear in Eq.\
(\ref{Collinear.gamma.g}), to represent the actual physical
suppression of the cross section at low $Q_1^2$. In fact, the Sudakov
form factor should not be used in this formula, since the raison
d'{\^e}tre of (\ref{Collinear.gamma.g}) is to be a subtraction term for
the NLO contribution to the cross section.  The unsubtracted NLO
contribution---Eq.\ (\ref{Unsubtracted.gamma.g}) below---has the same
singularity and lacks a Sudakov factor.  The subtraction will cancel
the singularity---see Eq.\ (\ref{Subtracted.gamma.g.BS})---to leave an
NLO term that is dominantly in the region of large $Q_1^2$.  Thus a
strict expansion to lowest order in $\alpha_s(Q^2)$ is appropriate, and a
resummation of higher-order terms, such as is represented by the
Sudakov form factor, is not needed.

\section{Bengtsson and Sj{\"o}strand algorithm}
\label{sec:correction}

Bengtsson and Sj{\"o}strand \cite{BS} give a prescription for the
4-momenta of the partons, and we now apply it to our first-order
calculation.  In this scheme, the 4-vectors for the momenta $q$, $p_1$
and $p_3$ of the virtual photon, the intermediate quark and the
incoming gluon obey the following requirements:
\begin{enumerate}
   
\item $q^\mu$ is the correct value of the photon's momentum.  

\item The proton is to be moving in the $-z$ direction.

\item $p_1^2 = -Q_1^2$.

\item $(p_1+q)^2 = (p_3-p_1)^2 = p_3^2 = 0$.

\item $z_1 = \displaystyle \frac{ p_1 \cdot (p_1+q) }
                                { p_3 \cdot (p_3+q) }
      $.

\end{enumerate}
In the $\gamma^*g$ center-of-mass frame, we therefore have
\begin{eqnarray}
\label{q.mu}
   q^\mu &=& \frac{Q ~ (1-Q_1^2/Q^2) } {2 \sqrt{ z_1 (1-z_1-Q_1^2/Q^2) } }
          \left( 1 - \frac{ 2z_1 }{ 1 - Q_1^2/Q^2 }, 
                 ~ \mathbf{0}_T,
                 ~ 1
          \right),
\\
\label{p.1prime.mu}
   p_{1'}^\mu &=&
      \frac{Q \sqrt{1 -z_1 - Q_1^2/Q^2} } 
           {2\sqrt{z_1} } 
\\
   && \times
      \left( 1, 
             ~ \frac{2Q_1}{Q}
                \frac{\sqrt{z_1 [1-(1+z_1)Q_1^2/Q^2]}}
                     {1 - Q_1^2/Q^2}
                \mathbf{n}_T, 
            ~ 1 - \frac{ 2z_1 Q_1^2/Q^2 }{ 1-Q_1^2/Q^2 }
      \right),
\nonumber
\\
\label{p.2.mu}
   p_2^\mu &=&
      \frac{Q \sqrt{1 -z_1 - Q_1^2/Q^2} } 
           {2\sqrt{z_1} }
\\
   && \times
      \left( 1, 
             ~ -\frac{2Q_1}{Q}
                \frac{\sqrt{z_1 [1-(1+z_1)Q_1^2/Q^2]}}
                     {1 - Q_1^2/Q^2}
                \mathbf{n}_T, 
            ~ -1 + \frac{ 2z_1 Q_1^2/Q^2 }{ 1-Q_1^2/Q^2 }
      \right),
\nonumber
\\
\label{p.3.mu}
   p_3^\mu &=&
      \frac{Q \, (1- Q_1^2/Q^2)} 
           {2\sqrt{ z_1 (1 - z_1 - Q_1^2/Q^2) } }
      ~ \left( 1, \, \mathbf{0}_T, \, -1 \right).
\end{eqnarray}
Here $\mathbf{n}_T$ is a unit transverse vector in the direction
defined by the azimuthal angle $\phi$.  The components are in the order
$(0,{\rm transverse}, z)$.

We now transform the cross section in Eq.\ (\ref{Collinear.gamma.g})
into convenient variables for a hard scattering describing
photon-gluon fusion.

The scattering angle obeys
\begin{eqnarray}
   \cos \theta &=& \frac{ p_1^z + q^z }{ p_1^0 + q^0 }
          = 1 - \frac{ 2z_1 Q_1^2/Q^2 }{ 1-Q_1^2/Q^2 } .
\end{eqnarray}
The fractional momentum of the gluon is defined in \cite{BS} to be
\begin{eqnarray}
\label{x3.original}
    x_3 &=& \frac{ p_3 \cdot (p_3+q) }{ p \cdot q }
      = \frac{x}{z_1} \left( 1 - \frac{Q_1^2}{Q^2} \right ) ,
\end{eqnarray}
where $p^\mu$ is the proton's momentum.  Up to a correction of order
$m^2/Q^2$, this definition agrees with the definition in terms of
light-front variables: $x_3 = ( p_3^0 - p_3^z )/( p^0-p^z )$,
since the gluon is on-shell with zero transverse momentum.  

We will also use the inverse transformation, to give $Q_1^2$ and $z_1$
in terms of $x_3$ and $\theta$:
\begin{eqnarray}
   \frac{Q_1^2}{Q^2} &=& (1 - \cos \theta ) \, \frac{x_3}{2x} ,
\nonumber\\
   z_1 &=& \frac{x}{x_3} - \frac{1}{2} \left(1 - \cos\theta \right) .
\end{eqnarray}
The Jacobian of the transformation is
\begin{equation}
   \frac{ \partial( x_3, \cos\theta ) }{ \partial( z_1, Q_1^2 ) }
  = \frac{ 2x }{ z_1 Q^2 },
\end{equation}
Then the cross section is: 
\begin{eqnarray}
\label{Collinear.gamma.g.BS1}
     \frac{ d\sigma^{\rm (BS1)}_{\rm shower} }
          { dx \, dy \, dx_3 \, d\!\cos\theta \, d\phi }
  &=& K  \sum_a e_a^2 \frac{\alpha_s(Q^2)}{4\pi^2}
       C(Q_1^2)
\\
  &&
       \frac{1}{1-\cos\theta}
       P\left( \frac{x}{x_3} - \frac{1}{2}(1-\cos\theta) \right) 
       \frac{x}{x_3} f_g(x/z_1,Q^2) .
\nonumber
\end{eqnarray}
The superscript on the cross section is `BS1' rather than `BS'.  This
anticipates that we will be forced to change the algorithm; it is the
modified cross section that will be denoted by a superscript `BS'.

\subsection{Inconsistency in kinematics}
\label{sec:inconsistency}

Observe that in Eq.\ (\ref{Collinear.gamma.g.BS1}) the fractional
momentum argument of the gluon density is $x/z_1$ rather than the
actual fractional momentum of the gluon, $x_3$. 

This is a symptom of the inconsistency explained by Bengtsson and
Sj{\"o}strand \cite{BS}.  The problem can be seen quite dramatically by
applying our first-order calculation to the case of a gluon target.
In that case the density of gluons is a delta-function:
$f_g(x_3,Q_1^2) = \delta(x_3-1)$, and the generated value of $z_1$ is 
$x$, from Eq.\ (\ref{Collinear.gamma.g}).  But then the definitions of the
parton 4-momenta imply that the fractional momentum of the gluon is 
$x_3 = 1-Q_1^2/Q^2$---see Eq.\ (\ref{x3.original})---instead of the
correct value of $x_3=1$.

\subsection{Parton model}

What has gone wrong can be explained by examining the derivation
of the parton-model formula.  We consider graphs for deep inelastic
scattering that have the form of Fig.\ \ref{parton.model}.  For our
purposes we can ignore polarization effects, so that the
contribution of this graph to the structure function is
\begin{equation}
\label{parton.model.graph}
   F = \frac{ Q^2 }{ 2\pi }
       \int \frac{ d^4p_1 }{ (2\pi)^4 }
       L(p, p_1) \, H(q^2, p_1^2, p_{1'}^2) \, U(p_{1'}^2) .
\end{equation}
Here $L$ represents the lower part of the graph, $H$ represents the
hard scattering, and $U$ represents the upper part of the graph.  The
overall factor $Q^2$ in the definition of the structure function
ensures that it obeys Bjorken scaling, and the overall factor $1/2\pi$
gives it the standard normalization

\FIGURE{
\centering
\psfrag{q}{$q$}
\psfrag{p_1}{$p_1$}
\psfrag{p_{1'}}{$p_{1'}$}
\psfrag{pp}{$p$}
\psfrag{U}{U}
\psfrag{L}{L}
\includegraphics[scale=0.45]{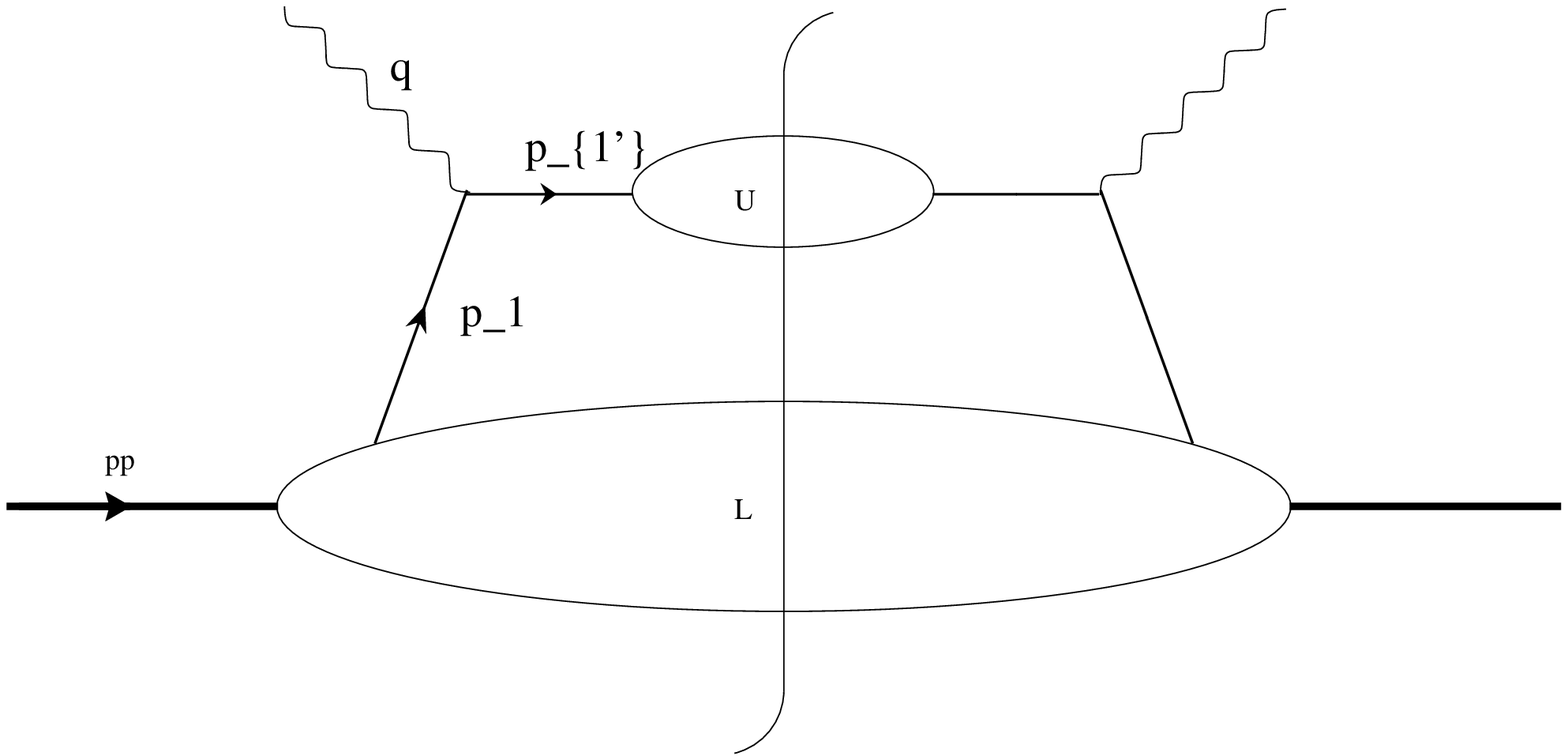}
\caption{The parton model for deep inelastic scattering is derived
   from cut graphs of this form.}
\label{parton.model}
}

We now use light-front coordinates [defined by 
$V^\mu = (V^+, V^-, \mathbf{v}_\perp)$, 
with $V^\pm = (V^0 \pm V^z) / \sqrt 2$]:
\begin{eqnarray}
   p^\mu     &=& \left( m^2/2p^-, \, p^-, \, \mathbf{0}_\perp \right), \\
   q^\mu     &=& \left( Q^2/2xp^-, \, -xp^-, \, \mathbf{0}_\perp \right), \\
   p_1^\mu   &=& \left( p_1^+, \, \xi p^-, \, \mathbf{p}_{1,\perp} \right), \\
   p_{1'}^\mu &=& 
      \left( 
       (Q^2+xm^2)/2xp^-, \, (\xi-x)p^-, \, \mathbf{p}_{1,\perp} 
      \right).
\end{eqnarray}

The parton-model approximation to the graph is obtained by making
approximations that are appropriate if the incoming and outgoing
quarks, $p_1$ and $p{_1'}$ have small transverse momenta and
virtualities, relative to $Q$.  To make this quantitative, we let the
magnitudes of $p_{1\perp}^2$, $p_1^2$, and $p_{1'}^2$ be $M^2$.  Then 
$\xi = x + O(M^2/Q^2)$.  Up to a power-suppressed correction, we can
replace $H$ by its value with massless quarks, $H(q^2,0,0)$, and we
can replace the value of $p_1^- = \xi p^-$ in $L$ by $xp^-$.  A shift in 
the integration over $\xi$ then gives the factorized form:
\begin{eqnarray}
   F &=& \left [ 
            xp^- \int \frac{ dp_1^+ \, d^2\mathbf{p}_{1,\perp} }{ (2\pi)^4 }
            L(p, \, (p_1^+,xp^-,\mathbf{p}_{1,\perp}))
       \right]
       H(q^2, 0, 0)
       \left[ \int \frac{ dp_{1'}^2 }{ 2\pi } \, dU(p_{1'}^2) \right]
\nonumber\\
     && + O(M^2/Q^2)
\nonumber\\
\label{parton.model.formula}
   &=& f(x) \, H(q^2, 0, 0) \, D  + O(M^2/Q^2) .
\end{eqnarray}
Here the parton density $f(x)$ is defined by the usual light-front
operator.  The fragmentation function $D$ is the integral over the
discontinuity of a propagator, so that it is equal to unity if the
integral is convergent.

The above derivation is exactly correct in a super-renormalizable
field theory, for then all the integrals over virtualities are
convergent in the approximated integral.  In a real theory, like QCD,
there are ultra-violet divergences that need to be renormalized.  The
correct physics is obtained in the factorization theorem which shows
that the above result is valid up to higher order corrections in
$\alpha_s(Q^2)$.

\subsection{Correction to the kinematics of initial-state showering}

In the parton-model formula (\ref{parton.model.formula}), the hard
scattering has been replaced by a massless approximation with zero
transverse momentum.  Moreover, in the usual way of estimating the
graph, the quark entering the fragmentation function is given zero
transverse momentum, and it is given a minus component of momentum
equal to $(\xi-x)p^-$.  For a calculation of the inclusive cross
section this use of approximated kinematics is correct, since the
hadronic final-state is integrated over.  But for the exclusive
calculation, as in an event generator, the use of approximated
kinematics is wrong, and it results in the inconsistent kinematics
noted in Sec.\ \ref{sec:inconsistency}.

The inconsistency is in the definitions of the variable
specifying the longitudinal momentum of the quark $p_1$.  Bengtsson
and Sj{\"o}strand \cite{BS} effectively\footnote{ 
   They actually define a variable $z$ for the relative longitudinal
   momenta of neighboring partons in the showering.  Their definition
   implies that the longitudinal momentum fraction of the struck quark 
   is given by Eq.\ (\ref{xprime}).  Note that the formula is not in
   its most satisfactory form.  It would appear more consistent to
   replace the denominator by $p\cdot(p+q)$.  This results in small
   changes, of relative size $xm^2/Q^2$.  To keep agreement with the
   formulae used in \cite{BS}, I do not make this further change.
}
define this variable by
\begin{equation}
\label{xprime}
  x_1 = \frac{ p_1 \cdot (p_1+q) }{ p \cdot q } .
\end{equation}
In contrast, the calculation of the Sudakov form factor in Eq.\
(\ref{Sudakov}) assumes that the fractional momentum of the struck
quark equals the Bjorken variable $x$, an assumption that is only
valid in the limit that transverse momenta, masses and virtualities
are negligible with respect to $Q$.

Hence the inconsistency arises because we have specified the
fractional momentum of $p_1$ in two ways.  On the one hand the
momentum $p_1^\mu$ is determined once one has specified the
virtualities of all the lines, the $z_1$ variable for the first
branching, and the fractional momentum of the incoming parton.  On the
other hand we asserted that the fractional momentum of $p_1$ is the
parton model value, i.e., that it equals the Bjorken variable $x$.
There are several ways one can correct the inconsistency:
\begin{itemize}

\item Change the value of the fractional momentum of $p_1$
   to the correct value.  This is the prescription of Bengtsson and
   Sj{\"o}strand \cite{BS}.  We will use it for the
   remainder of this section.  It is implemented in LEPTO and RAPGAP,
   and it requires a modified Sudakov form factor.

\item Only generate the value of Bjorken $x$ after the values of the
   fractional momenta are generated.  In this approach one cannot
   choose to generate events with a specified value of $x$. 

\item Remove the requirement on the fractional momentum of $p_1$,
   while keeping the unmodified Sudakov form factor. Here one
   generates the fractional momentum variables for all lines except
   $p_1$, and then determines $p_1$ without putting any explicit
   requirement on its fractional momentum.  This is the prescription I 
   will implement in Sec.\ \ref{sec:new.algorithm}.

\end{itemize}

I now explain the Bengtsson and Sj{\"o}strand method. They first make an
explicit definition equivalent to Eq.\ (\ref{xprime}):
\begin{equation}
\label{x1.prime}
  x_1 = x \left( 1 + \frac{ m_{1'}^2-Q_1^2 }{ Q^2 } \right) .
\end{equation}
Here, $m_{1'}^2 = p_{1'}^2$ and $Q_1^2 = -p_1^2$.  They then write a
modified Sudakov form factor:
\begin{equation}
\label{Sudakov.BS}
   S_a^{\rm (BS)}(x, Q_{\rm max}^2, Q_1^2, m_{1'}^2) 
   =
   \exp\left\{
      -\int_{Q_1^2}^{Q_{\rm max}^2} \frac{dQ'^2}{Q'^2}
       \frac{\alpha_s(Q'^2)}{2\pi}
       \sum_c \int_x^1 \frac{dz}{z}  P_{c\to ab}(z) 
                     \frac{f_c(x_1/z,Q'^2)}{f_a(x_1,Q'^2)}
   \right\} .
\end{equation}
At first sight, this modified Sudakov factor is not satisfactory,
since it depends on the virtuality of the outgoing jet.  This results
in a non-factorizing cross section, whereas one usually assumes that
the showering on different lines can be performed independently.  In
fact, there is no problem as regards the calculation.  One simply has
to perform the generation of the virtuality of the outgoing quark
before one performs the initial-state showering.  A practical
alternative is to use the original Sudakov factor to generate events
and then to reweight the events by the ratio of the form factors.  All
the necessary variables to define the modified Sudakov factor
(\ref{Sudakov.BS}) are available when the reweighting is done.  An
actual reweighting may be done, or the generated events may be vetoed
with the appropriate probability.

As we will see, the method also results in formulae for NLO cross
sections that are non-linear in the parton densities.  Similarly, the
relations between parton densities for the event generator and in the
\MSbar{} scheme are non-linear.  Although the complication in the
formulae must be regarded as a disadvantage, the disadvantage is not
fatal. 

A more fundamental concern is that the dependence of the Sudakov form
factor on the virtuality of the final-state parton may signal that the
form factor is not universal between different processes.  Since the
dependence is only on a single variable, one might hope that the
problem is tractable.

\subsection{Modification to first-order gluon-fusion}

The modification in the Sudakov factor entails two changes in the
cross section formula (\ref{Collinear.gamma.g.BS1}).  One is that
$f(x/z_1)$ must be replaced by $f(x_3)$, and the other is that a
factor $f_a(x)/f_a(x_1)$ must be inserted, where
$x_1 = x (1 - Q_1^2/Q^2)$.  Thus the first-order cross section
implemented in the event generator is
\begin{eqnarray}
\label{Collinear.gamma.g.BS}
     \frac{ d\sigma^{\rm (BS)}_{\rm shower} }
          { dx \, dy \, dx_3 \, d\!\cos\theta \, d\phi }
  &=& K  \sum_a e_a^2 \frac{\alpha_s(Q^2)}{4\pi^2}
       C(Q_1^2)
       \frac{x}{x_3} f_g(x_3,Q^2) 
       \frac{ f_a(x) }{ f_a(x_1) }
       \times
\nonumber
\\
  && \hspace*{1.3cm}
       \times \frac{1}{1-\cos\theta}
       ~ P\left( \frac{x}{x_3} - \frac{1}{2}(1-\cos\theta) \right) ,
\end{eqnarray}
where the fractional momentum of the gluon in the gluon density is now
the same as in the calculation of its 4-momentum.

The extra factor of $f_a(x)/f_a(x_1)$ looks quite unusual.  Nevertheless
it does represent what is actually implemented\footnote{
   Private communication from H. Jung and T. Sj{\"o}strand.
}
in the code for LEPTO and RAPGAP, following the prescription of \cite{BS}. 
This factor does go to unity when $Q_1^2 \ll Q^2$, as is necessary if
the showering is to be correct in the collinear limit.  However our
purpose is to derive the NLO correction to the event generator, and
for that we must use the actually implemented cross section when
$Q_1^2$ is of order $Q^2$.

\subsection{Photon-gluon fusion with subtraction}
\label{sec:NLO}

From standard references (e.g., \cite{text}), we find that the
unsubtracted photon-gluon fusion contribution to $F_2$ gives a cross
section
\begin{eqnarray}
\label{Unsubtracted.gamma.g}
     \frac{ d\sigma_{\rm unsubtracted}(\mbox{$F_2$ part}) }
          { dx \, dy \, dx_3 \, d\!\cos\theta \, d\phi }
  &=& K  \sum_{{\rm quarks}~a} e_a^2 \frac{\alpha_s(Q^2)}{4\pi^2}
       \frac{x}{x_3} f_g(x_3,Q^2) \times
\\
    && ~
    \times \left\{
       P(z) \left[ \frac{1}{1-\cos\theta} + \frac{1}{1+\cos\theta}
                  \right]
       -\frac12 + 3z(1-z)
    \right \} ,
\nonumber
\end{eqnarray}
where $z = x/x_3$. Notice that the sum is over quark, but not
antiquark flavors.  

We must now subtract the $O(\alpha_s)$ term associated with the showering, Eq.\
(\ref{Collinear.gamma.g.BS}), for both quarks and antiquarks.  The
antiquark term can be treated as a quark term with $\theta$ replaced by
$\pi-\theta$.  This gives
\begin{eqnarray}
\label{Subtracted.gamma.g.BS}
     \frac{ d\sigma^{\rm (BS)}_{\rm hard}(\mbox{$F_2$ part}) }
          { dx \, dy \, dx_3 \, d\!\cos\theta \, d\phi }
  &=& K  \sum_{{\rm quarks}~a} e_a^2 \frac{\alpha_s(Q^2)}{4\pi^2}
       \frac{x}{x_3} f_g(x_3,Q^2) \times
\nonumber\\
    && 
    \times \left\{
       \frac{ 1 }{ 1 - \cos\theta }
       \left[
          P(z) 
          ~ - ~ C(-t) 
            ~ \frac{ f_a(x) }{ f_a(x_{1t}) } ~
            P\left(z-\frac{1}{2}(1-\cos\theta)\right) 
       \right]
    \right.
\nonumber\\
    && ~
       +
       \frac{ 1 }{ 1 + \cos\theta }
       \left[
          P(z) 
          ~ - ~ C(-u) 
            ~ \frac{ f_{\bar a}(x) }{ f_{\bar a}(x_{1u}) } ~
            P\left(z-\frac{1}{2}(1+\cos\theta)\right) 
       \right]
\nonumber\\
    && ~
       -\frac12 + 3z(1-z)
    \Bigg \} .
\end{eqnarray}
Here again $z = x/x_3$, and
$P(z)=\frac12 (1-2z+2z^2)$.  
The virtualities in the cut-off function are
$-t = Q^2(1-\cos\theta) x_3 / 2x$ and 
$-u = Q^2(1+\cos\theta) x_3 / 2x$;
these are used to give $x_{1t} = x(1+t/Q^2)$ and $x_{1u} = x(1+u/Q^2)$.

Observe how the redefinition of the kinematics of the parton showers
has resulted in a formula for the NLO correction that is non-linear in 
the parton densities.  In principle this is not incorrect.  However,
the structure of the formula is much different to what one usual deals 
with, and is more complicated.  

For completeness, here follows the corresponding contribution to the
cross section that results from the photon-gluon fusion part of $F_L$:
\begin{equation}
\label{L.gamma.g}
     \frac{ d\sigma_{\rm hard}(\mbox{$F_L$ part}) }
          { dx \, dy \, dx_3 \, d\!\cos\theta \, d\phi }
  = - \frac{ K y^2 }
           { 2 - 2y + y^2 }
      \sum_{{\rm quarks}~a} e_a^2 \frac{\alpha_s(Q^2)}{4\pi^2}
       \frac{x}{x_3} f_g(x_3,Q^2) 
      \, 2z(1-z) .
\end{equation}
The complete gluon-fusion contribution to the cross section is the sum 
of (\ref{Subtracted.gamma.g.BS}) and (\ref{L.gamma.g}).

In the above calculations, there is an arbitrary cut-off function
$C(Q_1^2)$.  To implement an event generator with the standard Sudakov
form factor, one should set $C(Q_1^2) = \theta(Q^2-Q_1^2)$.  However, the
derivation of the leading order formalism does not require this
choice; the purpose of the derivation is only to obtain the correct
cross section when $Q_1^2 \ll Q^2$.  Another choice of the cut-off
function which is unity at small $Q_1^2$ would also be valid.

Now the aim of computing higher-order corrections, such as Eq.\
(\ref{Subtracted.gamma.g.BS}) is to obtain the correct cross section when
large virtualities are involved.  So any change in the cut-off
function is compensated by the corresponding changes in the
subtraction terms, up to errors of yet higher order in
$\alpha_s(Q^2)$.  So there is a kind of renormalization-group
invariance under changes in the cut-off function.  If we were able to
calculate the cross section to all orders in $\alpha_s$, then the
cross section would be exactly invariant under changes in $C(Q_1^2)$.

Although the cut-off function is arbitrary, it is not completely
arbitrary if we are to do useful calculations.  We must choose it so
that higher-order corrections, such as Eq.\
(\ref{Subtracted.gamma.g.BS}), are not excessively large.  Thus the
standard case $C(Q_1^2) = \theta(Q^2-Q_1^2)$ is a simple rational
choice.  However, if one examines the ranges of virtualities actually
involved, one may well find that some other choice is better.

Indeed, there is no need to restrict one's attention to sharp
cut-offs.  A smoother function, like 
\begin{equation}
\label{smooth.cut.off}
   C(Q_1^2) = \left \{
              \begin{array}{l@{~}l}
                 1 - A Q_1^2 / Q^2 & \mbox{if $Q_1^2 < Q^2 / A$}, \\[1mm]
                 0              & \mbox{otherwise}
              \end{array}
            \right. 
\end{equation}
might even be better, since then one could for example choose it to
track the typical behavior of a higher-order correction.  In this
formula $A$ is an adjustable parameter. 
{\em If the
cut-off function is changed, then one must also redefine the Sudakov
form factor.}  Such a change is likely to be particularly useful in
connection with the problem that Eq. (\ref{Subtracted.gamma.g.BS})
does not automatically give a positive cross section.

This will be discussed further in Sec.\ \ref{sec:negative-weights},
where I will show that the use of a suitable cut-off function can at
least substantially reduce the number of negative-weighted events, and 
where I will also discuss the extent how the use of a formula
that contains a negative cross section is not necessarily
incompatible with its use in an event generator.

\subsection{Comparison with \MSbarbold{} scheme}
\label{sec:scheme}

Once one has a systematic algorithm for treating NLO corrections in
event generators, it is both possible and necessary to answer the
question of what scheme is being used for the parton densities.  The
numerical values of the parton densities can be changed at order
$\alpha_s$ and beyond by a change of scheme.  Thus it is necessary to know
the actual scheme being used in order that the cross section is
actually known to the claimed accuracy of $\alpha_s$.  Moreover, standard
fits to parton densities are typically made in the \MSbar{} scheme; it
is necessary to translate these to the scheme used in the event
generator.

It does not seem to me to be possible to modify the Monte-Carlo
algorithm to use \MSbar{} densities directly, since the algorithm
explicitly uses the dependence on parton 4-momentum of the parton
correlation function in the target.  In contrast, the definition of
the \MSbar{} densities involves an integral over all values of the
transverse momentum and virtuality, followed by a renormalization of
the consequent ultra-violet divergence.   So one must resign oneself
to the fact that the event generator uses parton densities that are in
effect tailored to its algorithm.

A fundamental way of approaching this issue would be to deduce, from
derivation of the Monte-Carlo algorithm, the operators that define its
parton densities.  

But here I will take a lower-level approach, which is to integrate the
Monte-Carlo cross section over the hadronic final states and then to
require that the result be the same as the cross section in the
standard factorization approach with \MSbar{} parton densities.

In this approach one must be concerned that the relation between the
parton densities might be process dependent.  In fact, the process
dependence cancels.  For example, the same unsubtracted photon-gluon
cross-section Eq.\ (\ref{Unsubtracted.gamma.g}) occurs in both the
Monte-Carlo calculation and the \MSbar{} calculation, and so it
cancels out when one subtracts the two formulae for the same physical
cross section to obtain the relation between the parton densities. 

First we take the formula for the structure function with \MSbar{}
parton densities \cite{text}.  Now our aim is to compute a
process-independent relation between parton densities, and we want a
result for each flavor of quark and not just the combination that
appears in the usual electromagnetic $F_2$.  So it is convenient to
replace the actual cross section by one in which the photon couples
only to one flavor of quark, with unit coupling:
\begin{eqnarray}
\label{F2.MSbar}
  F^a_2(x,Q^2)
  &=&  x f^{(\MSbar)}_a(x, \mu^2)
\nonumber\\
  &&
       + \frac{ \alpha_s(\mu^2) }{ 2\pi }
       \int_x^1 dx_3 
       \frac{x}{x_3} f^{(\MSbar)}_g(x_3, \mu^2) 
 \left[
          P(z) \ln \frac{ Q^2 (1-z) }
                              { \mu^2 z }
          - \frac12 + 4z(1-z)
       \right]
\nonumber\\
   && + ~ \mbox{first-order quark terms} + O(\alpha_s^2),
\end{eqnarray}
where $z=x/x_3$, the same as in the formula for the hard scattering.
This must equal the same structure function given by the Monte-Carlo
calculation.  From the order $\alpha_s$ calculation in Eq.\
(\ref{Subtracted.gamma.g.BS}), we find 
\begin{eqnarray}
\label{F2.MC}
  F^a_2(x,Q^2)
  &=&  x f^{\rm (BS)}_a(x, Q^2)
\nonumber\\
    && 
       + ~ \frac{ \alpha_s(Q^2) }{ 2\pi }
       \int_x^1 dx_3 \int_{-1}^1 d\!\cos\theta
       \frac{x}{x_3} f_g(x_3,Q^2) 
       \times
\nonumber\\
    && ~~
    \left\{
       \frac{ 1 }{ 1 - \cos\theta }
       \left[
          P(z) 
          ~ - ~ C(-t) 
            ~ \frac{ f_a(x) }{ f_a(x_1) } ~
            P\left(z-\frac{1}{2}(1-\cos\theta)\right) 
       \right]
    \right.
\nonumber\\
    && ~~~
       -\frac14 + \frac32 z(1-z)
    \Bigg \} 
\nonumber\\
    && 
    + ~ \mbox{first-order quark terms} + O(\alpha_s^2).
\end{eqnarray}
Here $-t=Q^2 (1-\cos\theta) x_3 / 2x$, 
and $x_1 = x - \frac12 x_3 (1-\cos\theta)$.  Because of the
dependence on $f_a(x_1)$, it is not possible to do the integral over
$\cos\theta$ analytically.  The superscript on the quark density
indicates that it is in the scheme appropriate for the
Bengtsson-Sj{\"o}strand algorithm. 

It follows immediately that
\begin{eqnarray}
\label{pdf.relation.BS}
  x f^{\rm (BS)}_a(x, Q^2)
  &=&  x f^{(\MSbar)}_a(x, \mu^2)
+
\nonumber\\
   && 
\hspace*{-1cm}
       + \frac{ \alpha_s(\mu^2) }{ 2\pi }
       \int_x^1 dx_3 
       \frac{x}{x_3} f^{(\MSbar)}_g(x_3, \mu^2) 
\nonumber\\
   && 
       \times \Bigg\{
          P(z) \ln \left( \frac{ Q^2 (1-z) }
                              { \mu^2 z }
                   \right)
          + z(1-z)
-
\nonumber\\
   && 
~ ~ ~
       -  \int_{-1}^1 \frac{ d\!\cos\theta }{ 1 - \cos\theta }
       \left[
          P(z) 
          ~ - ~ C(-t) 
            ~ \frac{ f_a(x) }{ f_a(x_1) } ~
            P\left(z-\frac{1}{2}(1-\cos\theta)\right) 
       \right]
     \Bigg\}
\nonumber\\
   && 
\hspace*{-1cm}
   + ~ \mbox{first-order quark terms} + O(\alpha_s^2).
\end{eqnarray}
This is clearly a rather unpleasant formula.  However it is
necessitated by the algorithm currently used in the event generators.

\section{New algorithm for parton kinematics}
\label{sec:new.algorithm}

The formulae generated in the preceding sections are rather
complicated and have non-linear dependence on the parton densities.
This is a result of the particular choice made in \cite{BS} for
handling the kinematics of the off-shell struck quark.  So I now
propose an alternative algorithm which will result in much more
pleasant properties for the cross section and the parton densities.

\FIGURE{%
\centering
\psfrag{q}{$q$}
\psfrag{p_1}{$p_1$}
\psfrag{p_3}{$p_3$}
\psfrag{p_5}{$p_5$}
\includegraphics[scale=0.5]{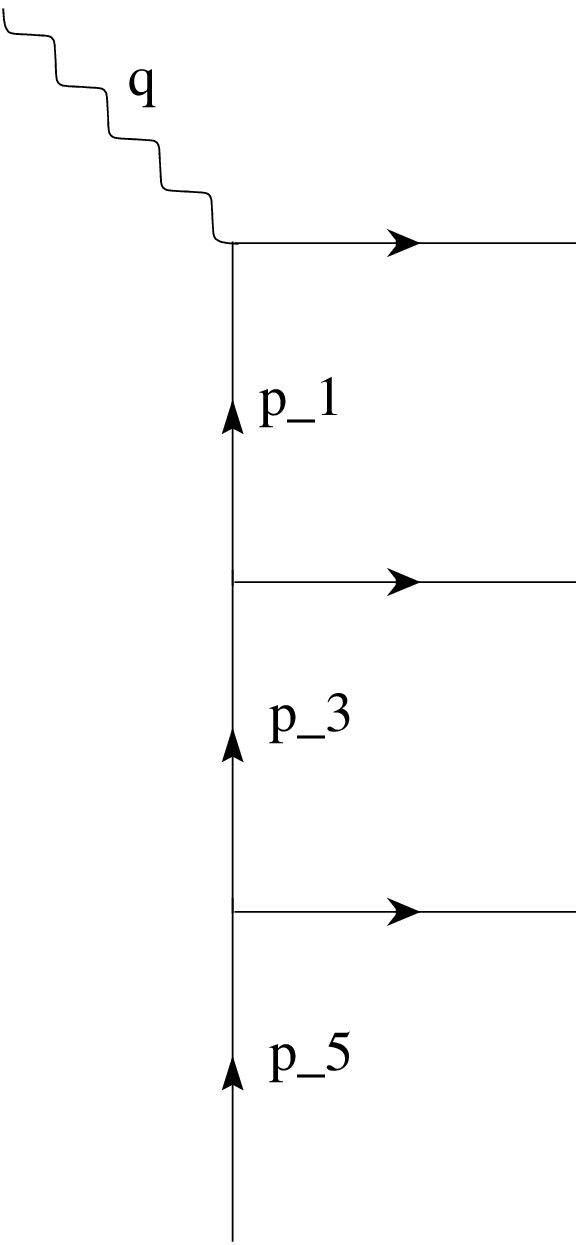}
\caption{Generic initial-state shower.}
\label{fig:shower}
}

Consider first how the 4-momenta of the partons are generated.  In
Fig.\ \ref{fig:shower} is shown a general shower.  The showering
algorithm generates a specification of the partons.  The specification 
consists of (a) the flavor of each parton, (b) the virtuality of each 
parton, (c) the fractional momentum $z_{2n+1}$ for each branching, and (d)
an azimuthal angle for the transverse momentum of each branching.  The 
fractional momentum of each space-like line is specified as
\begin{equation}
\label{x.from.z}
   x_{2n+1} = \frac{ x }{ \prod_{i=0}^{n-1} z_{2i+1}} ,
\end{equation}
so that $x_3 = x/z_1$, $x_5 = x/(z_1z_3)$, etc.  In reconstructing the 
4-momenta, Bengtsson and Sj{\"o}strand tell us to use
\begin{equation}
\label{p.from.x}
   \frac{ p_{2n+1} \cdot ( p_{2n+1} + q ) }{ p \cdot q } = x_{2n+1} .
\end{equation}
If we apply these equations for all $n$ from $0$ upwards, we get the
previously explained inconsistency.  The resolution proposed by
Bengtsson and Sj{\"o}strand was to change the formula, (\ref{x.from.z}); 
they made a consistent definition of a fractional momentum $x_{2n+1}$
as the right-hand side of Eq.\ (\ref{x.from.z}) times 
$[ 1 + ({m'}_1^2-Q_1^2)/Q^2 ]$, and they used the modified Sudakov
form factor (\ref{Sudakov.BS}).

An alternative which generates simpler formulae is as follows:
\begin{enumerate}

\item Generate virtualities for the partons and fractional momenta for 
   the branchings from the unmodified Sudakov form factor Eq.\
   (\ref{Sudakov}).  

\item Use Eq.\ (\ref{x.from.z}) to compute the fractional momenta,
   {\em except in the case $n=0$}, i.e., except for computing $x_1$. 

\item Similarly, use Eq.\ (\ref{p.from.x}) for all $p_{2n+1}$, except
   for $p_1$.

\item The momentum $p_1$ is now completely determined, given the
   virtualities and the azimuthal angles. 

\end{enumerate}
The inconsistency is eliminated because we have dropped one equation,
for the longitudinal momentum of $p_1$.  Bengtsson and Sj{\"o}strand, in
contrast, changed the computation of every $x_{2n+1}$ so that the two
conditions for $p_1$ are consistent.  

The price of the new procedure is that the definition of the
longitudinal momentum fraction, Eq.\ (\ref{p.from.x}) is not applied
universally.  The reward, as we will see, is that the formulae for the 
cross sections etc.\ are now of a conventional form that is linear in 
the parton densities and that is therefore considerably simpler.

Both procedures are correct; they simply represent alternative
prescriptions for computing the 4-momenta of the partons from the
scalar values generated by the algorithm.

\subsection{First-order term for showering}

Since the showers are generated from the unmodified Sudakov factor,
Eq.\ (\ref{Sudakov}), the cross section for the showering at order
$\alpha_s$ is given by the unmodified Eq.\ (\ref{Collinear.gamma.g}).
The parton momenta are given by simpler formulae than Eqs.\
(\ref{q.mu}--\ref{p.3.mu}):\footnote{
   In the list of requirements above Eq.\ (\ref{q.mu}), we only need
   to modify the formula for $z_1$, which now becomes
   $z_1 = Q^2 / 2 p_3\cdot (p_3 + q)$. 
}
\begin{eqnarray}
\label{q.mu.new}
   q^\mu &=& \frac{ Q }{ 2 \sqrt{\strut z_1 (1-z_1) } }
          \left( 1 - 2z_1, 
                 ~ \mathbf{0}_T,
                 ~ 1
          \right),
\\
\label{p.1prime.mu.new}
   p_{1'}^\mu &=&
      \frac{ Q \sqrt{\strut 1 -z_1} }{ 2 \sqrt{\strut z_1} } 
      \left( 1, ~ \mathbf{n}_T \sin\theta , 
            ~ \cos\theta
      \right),
\\
\label{p.2.mu.new}
   p_2^\mu &=&
      \frac{ Q \sqrt{\strut 1 -z_1} }{ 2 \sqrt{\strut z_1} } 
      \left( 1, ~ - \mathbf{n}_T \sin\theta,
            ~ -\cos\theta
      \right),
\\
\label{p.3.mu.new}
   p_3^\mu &=&
      \frac{ Q }{ 2\sqrt{ \strut z_1 (1 - z_1) } }
      ~ \left( 1, \, \mathbf{0}_T, \, -1 \right).
\end{eqnarray}
Here, we now have
\begin{equation}
\label{cos.theta.new}
   \cos\theta = 1 - \frac{ 2 z_1 Q_1^2 }{ Q^2 } ,
\end{equation}
so that
\begin{equation}
   \sin\theta = \frac{ 2 Q_1 }{ Q }
                \sqrt{ z_1 \left( 1 - z_1 \frac{Q_1^2}{Q^2} \right) } .
\end{equation}

The change of variables is now to $\cos\theta$ defined by Eq.\
(\ref{cos.theta.new}) and to $x_3 = x/z_1$.  This gives
\begin{equation}
\label{Collinear.gamma.g1.new}
     \frac{ d\sigma^{\rm (new)}_{\rm shower} }
          { dx \, dy \, dx_3 \, d\!\cos\theta \, d\phi }
  = K  \sum_a e_a^2 \frac{\alpha_s(Q^2)}{4\pi^2}
       \frac{ C(Q_1^2) }{ 1-\cos\theta }
       \, P\!\left( \frac{x}{x_3} \right) 
       \frac{x}{x_3} f_g(x_3,Q^2) .
\end{equation}

\subsection{NLO term with subtraction}

The hard scattering cross section with its subtraction is changed to
\begin{eqnarray}
\label{Subtracted.gamma.g.new}
     \frac{ d\sigma^{\rm (new)}_{\rm hard}(\mbox{$F_2$ part}) }
          { dx \, dy \, dx_3 \, d\!\cos\theta \, d\phi }
  &=& K  \sum_{{\rm quarks}~a} e_a^2 \frac{\alpha_s(Q^2)}{4\pi^2}
       \frac{x}{x_3} f_g(x_3,Q^2) 
\\
&& ~
    \left\{
       \frac{ P(z) [1 - C(-t)] }{ 1 - \cos\theta }
       + \frac{ P(z) [1 - C(-u)] }{ 1 + \cos\theta }
       -\frac12 + 3z(1-z)
    \right \} ,
\nonumber
\end{eqnarray}
where $z = x/x_3$, $-t = Q^2(1-\cos\theta) x_3 / 2x$ and
$-u = Q^2(1+\cos\theta) x_3 / 2x$.  Notice the
considerable simplification compared with the previous scheme.

\subsection{Relation to \MSbarbold{} parton densities}

From Eq.\ (\ref{Subtracted.gamma.g.new}) it follows that the structure
function $F_2^a$ is
\begin{eqnarray}
\label{F2.new}
  F^a_2(x,Q^2)
  &=&  x f^{\rm (new)}_a(x, Q^2)
\nonumber\\
  &&
       + \frac{ \alpha_s(Q^2) }{ 2\pi }
       \int_x^1 dx_3 
       \frac{x}{x_3} f^{\rm (new)}_g(x_3, Q^2)
        \left[
          P(z) \ln (1/z)
          - \frac12 + 3z(1-z)
       \right]
\nonumber\\
   && + ~ \mbox{first-order quark terms} + O(\alpha_s^2),
\end{eqnarray}
where again $z=x/x_3$.
Comparison with the structure function expressed in terms of the
\MSbar{} densities, Eq.\ (\ref{F2.MSbar}), gives
\begin{eqnarray}
\label{pdf.relation.new}
  x f^{\rm (new)}_a(x, Q^2)
  &=&  x f^{(\MSbar)}_a(x, \mu^2)
\nonumber\\
   && + \frac{ \alpha_s(\mu^2) }{ 2\pi }
       \int_x^1 dx_3 
       \frac{x}{x_3} f^{(\MSbar)}_g(x_3, \mu^2) 
        \left[
          P(z) \ln \frac{ Q^2 (1-z) }{ \mu^2 }
          + z (1-z)
       \right]
\nonumber\\
   && 
   + \mbox{first-order quark terms} + O(\alpha_s^2).
\end{eqnarray}
This is clearly much more tractable than the previous version.

\section{Negative weighted events}
\label{sec:negative-weights}

The subtracted NLO cross sections Eqs.\ (\ref{Subtracted.gamma.g.BS})
and (\ref{Subtracted.gamma.g.new}) are not automatically positive.
This is not unphysical, since these formulae only give a component of
the cross-section.  However it is a problem for an event generator
when one generates separate classes of LO and NLO events.  

The problem is not necessarily fatal.  For example, suppose one wishes
to make a histogram of a differential cross section.  Then one can
choose to generate weighted events (which if necessary can be passed
through a detector simulation, etc).  The contents of each bin of the
histogram are obtained as a weighted sum of events, and there is no
need to require that individual events have positive weights.

What would cause a real disaster would be to have a relatively small
final answer being obtained by a cancellation of large numbers of
positive-weighted events and large numbers of negative-weighted
events.  This is not the case for the algorithm proposed in this
paper, but it is the case for some kinds of analytic calculation. 

Nevertheless, one often prefers to be able to generate unweighted
events, for then one can genuinely simulate real experimental data.
This can be done on the basis of generation of weighted events, by
standard Monte-Carlo techniques, but only if all the weights are
positive. 

Luckily there is freedom in the proposed method to reduce the number
of negative-weighted events, by the choice of the cut-off function
$C(Q_1^2)$.  Not only can the position of the cut-off be changed from
its standard value $Q^2$, but the shape of the function can be
changed, for example to the form in Eq.\ (\ref{smooth.cut.off}).
Appropriate changes in the showering algorithm will be needed, as
explained in Sec.\ \ref{sec:NLO}.  At low $Q_1^2$, the cut-off
function goes to unity, so that collinear limits are unchanged.  But
the smoother form of cut-off reduces the amount that is subtracted
from the bare cross section at large $t$ or $u$ in Eqs.\
(\ref{Subtracted.gamma.g.BS}) and (\ref{Subtracted.gamma.g.new}).  For
a suitable cut-off function, the result is therefore to reduce the
number of negative-weighted events, if not eliminate them completely.

Even if one cannot eliminate absolutely all the negative-weighted
events, it should be enough for practical purposes if only a small
fraction of the generated events have negative weights.

\section{Future work}
\label{sec:future}

In the context of diffractive deep-inelastic scattering, the NLO
corrections calculated in this paper are clearly the most urgently
needed.  Because the gluon density is substantially larger than the
quark densities, these particular corrections are not numerically
suppressed in the way that would otherwise be expected of NLO
corrections.  The first new result is the calculation of the
gluon-induced contribution to the hard scattering [Eqs.\
(\ref{Subtracted.gamma.g.BS}) and (\ref{Subtracted.gamma.g.new})].
The second new result is the gluon-induced part of the relation
between the parton densities to be used in the event generator and the
\MSbar{} parton densities [Eqs.\ (\ref{pdf.relation.BS}) and
(\ref{pdf.relation.new})].

We saw that within the conventional scheme for performing
initial-state showering, the NLO corrections are unusually complicated 
non-linear functionals of the parton densities.  This is an inevitable
consequence of the algorithm chosen in \cite{BS} for imposing
consistent parton kinematics.  Therefore, in Sec.\
\ref{sec:new.algorithm}, I proposed an alternative algorithm for
computing the parton 4-momenta.  The change in algorithm entails a
change in the scheme that defines the parton densities used in the
event generator.  Different numerical values of the parton densities
are to be used in each scheme, and these numerical values differ from
the \MSbar{} parton densities that are obtained in the standard global
analyses.

Further calculations and theoretical developments that are needed
include: 
\begin{enumerate}

\item  Calculation at order $\alpha_s$ of the quark-induced subprocesses. 

\item Numerical calculations of the parton densities from the standard
   fits, which use the \MSbar{} scheme. 

\item Extension of the cross section calculations to include weak
   boson exchange.

\item Extension of the methods to handle other processes in
   hadron-hadron collisions and $e^+e^-$ annihilation.

\item Non-leading corrections to the showering.

\item Investigation of the non-positivity of the NLO term in the 
   cross sections:
   Can it be eliminated by a suitable choice of the cut-off function?
   How acceptable is it if an NLO term is negative?

\item Handling more general processes will require a proper treatment
   of the soft region.  Technically, I see this as the most difficult
   problem. 

\item The NLO corrections are valid when the virtuality of the
   intermediate quark, $Q_1^2$, is of order $Q^2$.  There is no
   singularity at $Q_1^2=0$, so the contribution from $Q_1^2 \ll Q^2$
   is suppressed by a power of $Q^2$.  Even so, events are generated
   for all values of $Q_1^2$, down to $Q_1^2=0$.  Thus one may need to
   modify the Monte-Carlo algorithm to handle this kinematic region.

\item One of the treatments of the kinematics of off-shell partons
   entails a redefinition of the Sudakov form factor in a
   way that appears non-universal.  It needs to be understood whether
   this non-universality is genuine or whether it is only apparent.

\item It would be useful to understand the relation of the Monte-Carlo
   formalism to definitions of the parton densities that involve
   explicit parton transverse momentum.  Compare Mrenna's work
   \cite{mrenna}. 

\item Would it be useful to find a way of computing the hard
   scattering coefficients with off-shell matrix elements?  There are
   obviously non-trivial issues of gauge invariance that would arise. 

\item Which of the algorithms for calculating the parton 4-momenta is
   better?  Is there a yet better algorithm?

\end{enumerate}

\acknowledgments

This work was supported in part by the U.S.\ Department of Energy
under grant number DE-FG02-90ER-40577.
I would like to thank H. Jung for patiently explaining the details 
of the algorithms, 
and I would also like to thank Y. Chen, S. Mrenna, T. Sj{\"o}strand and N.
Tkachuk for comments on drafts of this paper.
I thank S. Schilling for pointing out an error in the originally
published version of this paper.


\end{document}